\documentclass{aastex701}

\usepackage{adjustbox}
\usepackage{rotating}
\usepackage{makecell}
\usepackage{multirow}
\usepackage{booktabs}
\usepackage{amsmath}
\usepackage{float}
\usepackage{soul}

\begin{document}

\title{BSN-VII: Photometric Light Curve Study of Three Cool, Large-Amplitude, Low Mass,\\
W-Subtype Contact Binaries}

\author[0000-0002-0196-9732]{Atila Poro}
\altaffiliation{atila.poro@obspm.fr}
\affiliation{LUX, Observatoire de Paris, CNRS, PSL, 61 Avenue de l'Observatoire, 75014 Paris, France}
\affiliation{Astronomy Department, Raderon AI Lab., BC., Burnaby V5C 0J3, Canada}
\email{atila.poro@obspm.fr}

\author[0000-0003-1263-808X]{Raul Michel}
\affiliation{Instituto de Astronom\'ia, UNAM. A.P. 106, 22800 Ensenada, BC, M\'exico}
\email{rmm@astro.unam.mx}

\author{Diana Zadkhosh}
\affiliation{Farzanegan Institute, Khorramabad, Iran}
\email{dianazadkhosh@gmail.com}

\author{Ava Ekrami Fard}
\affiliation{Farhangian Institute, Khorramabad, Iran}
\email{ava.ekramifard@gmail.com}

\author{Kian Farhadi Sangdehi}
\affiliation{Damghan Branch of the National Organization for Development of Exceptional Talents (SAMPAD), Damghan, Iran}
\email{kian.f12403@gmail.com}

\author[0009-0007-2048-4865]{Anica Lekic}
\affiliation{IPSA Institut Polytechnique des Sciences Avancées, 94200 Ivry-sur-Seine, France}
\email{anica.lekic@ipsa.fr}

\author{Maxime Cellier}
\affiliation{IPSA Institut Polytechnique des Sciences Avancées, 94200 Ivry-sur-Seine, France}
\email{maxime.cellier@ipsa.fr}

\author{Chloé Da Graca Baptista}
\affiliation{IPSA Institut Polytechnique des Sciences Avancées, 94200 Ivry-sur-Seine, France}
\email{chloe.da-graca-baptista@ipsa.fr}

\author{Paul Pointier}
\affiliation{IPSA Institut Polytechnique des Sciences Avancées, 94200 Ivry-sur-Seine, France}
\email{paul.pointier@ipsa.fr}

\author[0000-0002-1972-8400]{Fahri Alicavus}
\affiliation{Çanakkale Onsekiz Mart University, Faculty of Science, Department of Physics, 17020, Çanakkale, Türkiye}
\affiliation{Çanakkale Onsekiz Mart University, Astrophysics Research Center and Ulupnar Observatory, 17020, Çanakkale, Türkiye}
\email{fahrilcvs@gmail.com}

\author[0009-0006-1033-5885]{Elham Sarvari}
\affiliation{BSN Project; Independent researcher, 12101 Berlin, Germany}
\email{fh.elhamsarvari94@gmail.com}

\author[0000-0001-9809-7493]{Neslihan Alan}
\affiliation{Fatih Sultan Mehmet Vakif University, Department of History of Science, 34664 Istanbul, Türkiye}
\email{neslihan.alan@gmail.com}


\begin{abstract}
We present a comprehensive photometric study of three W UMa-type contact binary systems, V1104 Her, V1284 Her, and V2822 Ori, based on a combination of ground-based observations and space-based TESS photometry. The light curves were modeled using the BSN application under a contact configuration, with parameters and their uncertainties estimated through the Markov Chain Monte Carlo (MCMC) approach. The derived solutions indicate that the target systems are in a shallow-contact configuration. The target systems are classified as W-subtype contact binaries, with the more massive component cooler than its companion. The light curve of V1284 Her shows a pronounced O'Connell effect, which is reproduced by introducing a cool starspot on the primary component. Absolute parameters were estimated using the empirical parameter relationship between orbital period and semi-major axis, yielding component masses in the range $0.48$-$0.86\,M_\odot$ and confirming the low-mass nature of the systems. Orbital period variations were investigated using eclipse timings from the literature, ground-based observations, and TESS data by examining both linear and cyclic models. Statistical model comparison indicates that the cyclic model provides a better description of the eclipse timings for V1104 Her and V1284 Her, whereas the linear model is preferred for V2822 Ori, although the difference between the two models is relatively small.
\end{abstract}


\section{Introduction}
W UMa-type contact binary systems constitute a distinct type of short-period eclipsing binaries whose physical characteristics are strongly shaped by compact orbital configurations and continuous interaction between the stellar components (\citealt{1968ApJ...153..877L,1978ASSL...68.....K,2005ApJ...629.1055Y}). Their physical state is governed by a delicate balance between gravitational, thermal, and rotational effects, making the derived physical parameters highly sensitive to small variations in the adopted system parameters and, consequently, to the details of the modeling (\citealt{2009MNRAS.397..857S,2009MNRAS.396.2176J}). While modern modeling techniques can achieve good agreement with the photometric light curve shapes, the resulting solutions are often affected by strong interdependencies among model parameters (\citealt{2026NewA..12302484L}). Inclination, mass ratio, degree of contact, fillout factor, and surface inhomogeneities can influence the light curve in similar ways, so small changes in one parameter may be compensated by others (\citealt{2023A&A...672A.176P}). This leads to multiple parameter combinations producing comparably good fits, highlighting the challenge of identifying a unique solution from photometric data alone. Consequently, photometric analyses may admit multiple physically plausible configurations, emphasizing the need to interpret model results within a broader physical context rather than relying solely on statistical fit quality.

In contact binary systems, the two stars share a common envelope that surrounds both components and facilitates continuous interaction. This shared envelope governs the distribution of energy and facilitates the process of thermal homogenization, which tends to reduce temperature differences between the stellar components in contact binaries (\citealt{2009MNRAS.396.2176J}). Hydrodynamic studies have shown that the temperature differences between the stellar components strongly depend on the rate and manner of energy exchange within this envelope (\citealt{2023A&A...672A.175F}). The neck region, where the stars are connected near the inner Lagrangian point, plays a crucial role in this process, acting as the main channel for energy flow between the components (\citealt{2011A&A...529A..11S}). Models indicate that large-scale circulations within the envelope redistribute heat efficiently, further reducing temperature contrasts (\citealt{2009MNRAS.397..857S}).

Large-amplitude contact binaries constitute an important subgroup of W UMa stars because their pronounced eclipse morphology provides stronger constraints on the system geometry, making them particularly valuable for investigating the mass ratio, degree of contact, and evolutionary status of contact binary systems. Recent observational studies have combined photometric and spectroscopic data to improve the accuracy of derived physical parameters for such systems. For example, \cite{2025AJ....169...85X} presented a detailed photometric and spectroscopic investigation of three large-amplitude contact binaries, while \cite{2026ApJS..284...65L} derived physical parameters for a large sample of contact binaries using combined spectroscopic and photometric observations, highlighting the importance of complementary datasets for constraining fundamental system properties.

After deriving the acceptable solutions from photometric light curves, the absolute parameters of the system, such as masses, radii, and luminosities, can be estimated using direct methods, Gaia DR3 parallaxes, or empirical parameter relationships (\citealt{2024NewA..11002227P,2025MNRAS.538.1427P}). This estimation allows the systems to be compared within a broader stellar context, with the resulting absolute parameters representing model-dependent estimates rather than direct measurements. Additionally, contact binary systems are commonly classified into W- and A-subtypes based on the temperatures and masses of their component stars, which are derived from light curve modeling and the estimation of absolute parameters (\citealt{1970VA.....12..217B}). In A-subtype systems, the more massive star is also the hotter component, whereas in W-subtype systems, the less massive star exhibits a higher surface temperature. This classification reflects the underlying mass–temperature distribution within the binary and provides important context for interpreting their evolutionary state and energy transfer processes.

Orbital period behavior offers an independent diagnostic of the physical processes shaping contact binary systems (\citealt{2021ApJS..254...10L,2024RAA....24a5002P}). Variations in the orbital period may reflect mass redistribution between the components, Angular Momentum Loss (AML) and energy dissipation through magnetic braking, or other dynamical influences operating within the system (\citealt{1992ApJ...385..621A,2004MNRAS.355.1383L}). Importantly, the orbital period carries information related to evolutionary processes acting over extended timescales that are not directly accessible through instantaneous photometric configurations (\citealt{2025PASP..137k4203P}). Nevertheless, patterns observed in period variation analyses often remain open to multiple interpretations, as different physical mechanisms can generate comparable timing signatures when considered in isolation. Comprehensive understanding of contact binaries benefits from a framework that considers photometric structure and orbital evolution simultaneously. Examination of light curve solutions alongside orbital period behavior enables consistency checks between inferred system geometry, derived physical parameters, and plausible evolutionary scenarios (\citealt{1971ApJ...166..605W,2006AJ....131.2986P}). Although such an approach does not remove degeneracies entirely, it provides a more balanced and physically grounded interpretation by constraining conclusions through complementary diagnostics (\citealt{2026NewA..12302484L}). This integrated perspective is particularly useful for systems approaching critical evolutionary or dynamical stages, as it allows cross-checking photometric and orbital period analyses to identify subtle but physically meaningful effects, such as changes in mass distribution, orbital period variations, or surface temperature differences.

Following the framework started by \cite{2025MNRAS.537.3160P}, this work expands the sample by incorporating new observations and a detailed analysis of additional W~UMa-type contact binary systems from the BSN\footnote{\url{https://bsnp.info/}}. In this work, we present a combined photometric and orbital period study of three W UMa-type contact binary systems using high-quality photometric data. The selected systems V1284 Her and V2822 Ori have not been previously analyzed, whereas V1104 Her was studied by \cite{2015AJ....149..148L} and \cite{loukaidou2022cobitom}. We examine the light curve characteristics, estimate absolute parameters, and investigate orbital period behavior within a unified analytical framework, aiming to provide a clearer and more coherent characterization of these short-period contact binary systems.

\vspace{0.6cm}
\section{Target Systems and Observation}
This study presents a detailed photometric analysis of three contact binary systems: V1104 Her (TIC 298661555), V1284 Her (TIC 157448336), and V2822 Ori (TIC 12312867). Table \ref{tab:systemsinfo} presents the astrometric and photometric properties of the target systems derived from Gaia DR3 data (\citealt{2023A&A...674A..33G}), together with effective temperatures and TESS-band magnitudes reported from the Transiting Exoplanet Survey Satellite (TESS) Input Catalog (TIC) v8.2. It should be noted that the distances presented in Table \ref{tab:systemsinfo} were directly adopted from the catalog by \cite{2021AJ....161..147B}, which provides estimates derived from Gaia DR3 parallaxes using the Bayesian method. Additionally, Table \ref{tab:systemsinfo} presents the maximum apparent magnitude of each target system in the $V$ filter ($V_{\mathrm max}$) derived from our observations.
Among the studied systems, V1104 Her and V1284 Her were detected by the Robotic Optical Transient Search Experiment I (ROTSE-I) telescope (\citealt{2000AJ....119.1901A}), while V2822 Ori was reported in the Gaia DR3 catalog of eclipsing binary candidates (\citealt{2023A&A...674A..16M}). The target systems are classified as contact binaries in several catalogs and databases, including the All-Sky Automated Survey for SuperNovae (ASAS-SN; \citealt{2018MNRAS.477.3145J}) and the Variable Star Index (VSX\footnote{\url{https://vsx.aavso.org/}}). Based on ASAS-SN reports, the target systems are classified as large-amplitude contact binaries, with reported photometric amplitudes of 0.89, 0.87, and 0.67 magnitude for V1104 Her, V1284 Her, and V2822 Ori, respectively. Additionally, all systems have effective temperatures below 4800 K (Table \ref{tab:systemsinfo}), classifying them as cool contact binaries (\citealt{2011mast.conf...86S}).
\\
\\
Ground-based photometry of V1104 Her and V1284 Her was carried out at the San Pedro Mártir Observatory in México, located at $31^\circ02^{'}39^{''}$ N, $115^\circ27^{'}49^{''}$ W, with an altitude of 2830\,m. V1104 Her and V1284 Her were observed on 2025 June 16 and 2025 June 14, respectively. Observations were performed using the 0.84\,m Ritchey-Chrétien telescope ($f/15$) equipped with a Marconi-5 CCD camera (Spectral Instruments) featuring an e2v CCD231-42 detector with $15\times15\,\mu\mathrm{m}^2$ pixels, a gain of $2.2\ e^-\ \mathrm{ADU}^{-1}$, and a readout noise of $3.6\ e^-$. Standard Johnson-Cousins $BVR_cI_c$ filters were used. Exposure times of $B$=90~s, $V$=50~s, $R_c$=35~s, and $I_c$=30~s were used for both V1104 Her and V1284 Her. All images were processed following conventional procedures in IRAF, including bias subtraction and flat-fielding, following the methodology described by \cite{1986SPIE..627..733T}.

Observations of the eclipsing binary V2822 Ori were obtained at the El Sauce Observatory in Chile, located at $30.472529^\circ$ S, $70.762999^\circ$ W, at an elevation of 1525\,m. The system was observed on four nights, namely 2024 November 28, November 30, December 13, and December 19. Data were collected using the CHI-1-CMOS telescope with a 610\,mm aperture and a focal length of 3962\,mm ($f/6.5$), equipped with a QHY~600M~Pro CMOS camera providing a $31 \times 20.7$ arcmin field of view and a pixel scale of 0.39\,arcsec\,pixel$^{-1}$. Data acquisition was performed through the $r'$ filter, and each image was exposed for 60 s. Photometric measurements and image reduction were carried out using the Muniwin\footnote{\url{http://c-munipack.sourceforge.net/}} software package (\citealt{motl2011muniwin}).
\\
\\
TESS was launched by NASA in 2018 with the goal of detecting exoplanets across the sky (\citealt{2010AAS...21545006R,2014SPIE.9143E..20R,2018AJ....156..102S}). TESS carries four wide-field cameras that sequentially monitor different sectors of the sky, with each sector observed for approximately 27.4 days. For the present study, we employed TESS time-series photometry of the three binary systems. The resulting light curves are obtained in the TESS “T” passband, which spans a broad wavelength range from 600 to 1000~nm (\citealt{2015JATIS...1a4003R}). The TESS time-series data used in this study, including the observed sectors and exposure lengths, are presented in Table \ref{tab:tess}. The TESS observations were retrieved from the Mikulski Archive for Space Telescopes (MAST)\footnote{\url{https://mast.stsci.edu/portal/Mashup/Clients/Mast/Portal.html}}. Light curves were extracted using the Lightkurve software package, and systematic trends were corrected according to the procedures implemented in the TESS Science Processing Operations Center (SPOC) pipeline (\citealt{2016SPIE.9913E..3EJ}).

\begin{table*}
\renewcommand\arraystretch{1.1}
\caption{Astrometric and photometric properties of the target systems.}
\centering
\begin{center}
\footnotesize
\begin{tabular}{c|ccccc|cc|c}
\hline
System & \multicolumn{5}{c|}{Gaia DR3} & \multicolumn{2}{c|}{TIC} & This study\\
& RA$.^\circ$(J2000) & Dec$.^\circ$(J2000) & $d$(pc) & RUWE & $T$(K) & $T$(K) & TESS mag. & $V_{\mathrm max}$(mag.)\\
\hline
V1104 Her & 272.448111 & 49.048502 & 184 & 1.068 & 4232(6) & 3975(130) & 12.60(3) & 13.43(10) \\
V1284 Her & 264.855464 & 36.783083 & 269 & 1.047 & - & 4709(180) &
12.18(3) & 12.66(9) \\
V2822 Ori & 92.065178 & -1.531041 & 122 & 1.080 & - & 4769(209) &
10.87(4) & 11.92(12) \\
\hline
\end{tabular}
\end{center}
\label{tab:systemsinfo}
\end{table*}

\begin{table*}
\renewcommand\arraystretch{1.1}
\caption{Used TESS time-series data in this investigation.}
\centering
\begin{center}
\footnotesize
\begin{tabular}{c c c}
\hline
System & TIC Id & Sector/Exposure length(s)\\
\hline
V1104 Her & 298661555 & 14/1800 - 25,26/120 - 41,52,54/600 - 58,74,79,81/200\\
V1284 Her & 157448336 & 25,26/1800 - 52,53/600 - 79,80/200\\
V2822 Ori & 12312867 & 33/600 - 87/200\\
\hline
\end{tabular}
\end{center}
\label{tab:tess}
\end{table*}

\vspace{0.6cm}
\section{Light Curve Modeling}
Photometric light curves of the target systems were analyzed using the BSN application (version 1.0; \citealt{2025Galax..13...74P}), which was specifically developed for modeling contact binary systems. According to \citealt{2025Galax..13...74P}, this program provides an enhanced environment for modeling by integrating a broader set of diagnostic tools, offering improved numerical stability, and supporting a user-friendly workflow suitable for both initial and advanced light-curve solutions. Streamlined parameter adjustment, visualization, and iterative refinement are facilitated through the interface, and the current version is available for the Windows operating system.

\begin{figure*}
\centering
\includegraphics[width=0.99\textwidth]{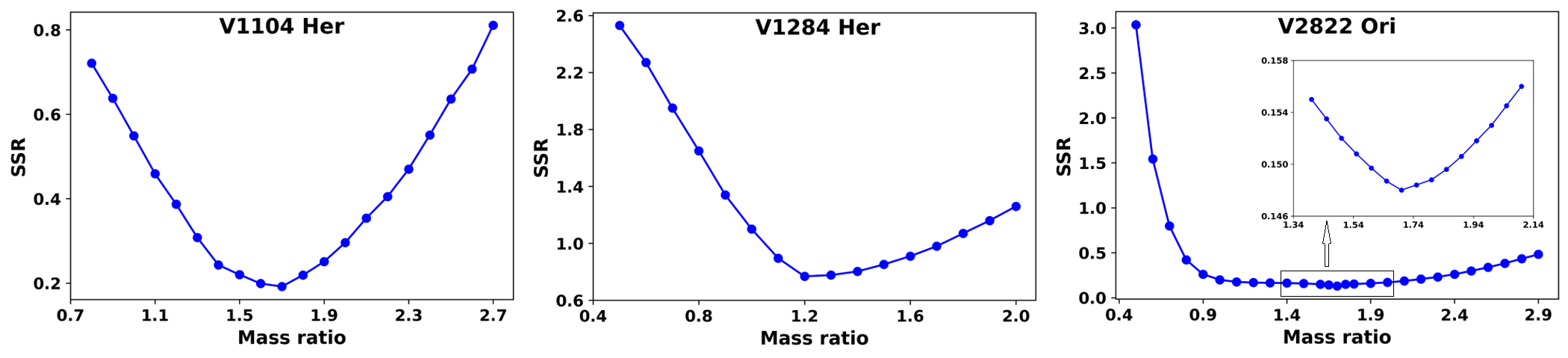}
\includegraphics[width=0.99\textwidth]{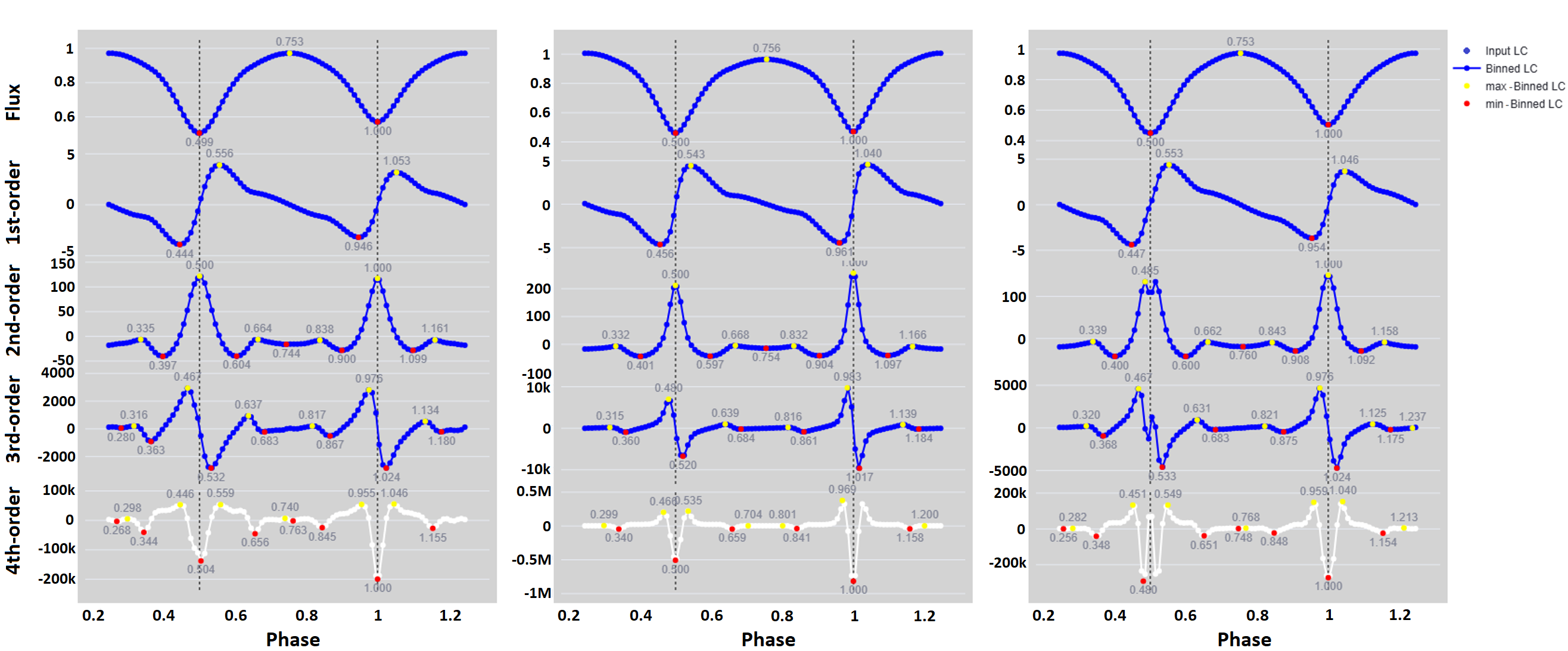}
\caption{Upper panel: Sum of squared residuals as a function of the mass ratio.
Lower panels: Observed photometric light curves and their first through third time derivatives (ordered from top to bottom), employed to estimate the initial mass ratio for each target system using the Kouzuma method. The vertical-axis units correspond to W~m$^{-2}$, 10~W~m$^{-2}$~day$^{-1}$, $10^{2}$~W~m$^{-2}$~day$^{-2}$, $10^{4}$~W~m$^{-2}$~day$^{-3}$, and $10^{6}$~W~m$^{-2}$~day$^{-4}$, respectively.}
\label{Fig:q}
\end{figure*}

Photometric modeling of the target systems was performed under a unified set of physical assumptions and modeling constraints. The contact configuration was adopted throughout, as the catalog classifications and short orbital periods of the target systems collectively indicate that the stellar components are in physical and thermal contact. Orbital phases were calculated from the observational times using the ephemerides described in Section~5. The gravity-darkening parameters were held fixed at $g_{1}=g_{2}=0.32$, consistent with \cite{1967ZA.....65...89L}, while the bolometric albedo values were adopted as $A_{1}=A_{2}=0.5$ following \cite{1969AcA....19..245R}. The atmospheric model presented by \cite{2004A&A...419..725C} was employed to represent the radiative properties of the individual stellar components. We initially intended to adopt the effective temperatures ($T$) of the target systems from the Gaia DR3 catalog; however, only V1104 Her has a reported temperature in this release (Table \ref{tab:systemsinfo}). Consequently, the effective temperatures for the remaining two systems were obtained from version 8.2 of the TESS Input Catalog (TIC). It should be noted that the reported uncertainties for temperatures in Gaia DR3 and TIC were not incorporated into the initial values used for modeling (\citealt{2025MNRAS.537.3160P}). The effective temperatures reported in Gaia DR3 and the TIC were taken to represent the hotter components of each system, as inferred from the relative depths of the light curve minima. The temperature of the cooler component was subsequently estimated from the depth difference between the primary and secondary minima. Another important parameter that needs to be defined at the outset of modeling is the mass ratio ($q$). Two distinct approaches were used to obtain initial estimates of the mass ratio for initiating the light curve analysis. The first approach involved applying the $q$-search method (\cite{2005Ap&SS.296..221T}). An initial broad range of $q$ values, from 0.05 to 20, was considered for all target systems. Once approximate minima were identified, a narrower range was examined to refine the estimates by minimizing the sum of squared differences between the observed and synthetic light curves (Table \ref{tab-solutions}). The second approach, introduced by \cite{2023ApJ...958...84K, 2025PASJ...77.1323K}, estimates the photometric mass ratio of overcontact binaries through analysis of higher-order derivatives of their light curves. This technique, unlike iterative methods such as the $q$-search or MCMC optimization, leverages characteristic features in the second- and third-order derivatives to infer the mass ratio. Implementation involves computing the third-order derivative, locating local extrema near the eclipse phases, and combining these with the orbital period to derive a parameter $W$, which exhibits a strong correlation with $q$. Validation against systems with known spectroscopic mass ratios shows that approximately 67\% of photometric estimates fall within the uncertainties, with 95\% deviating by less than $\pm0.1$. The method requires clearly defined maxima and minima in the relevant derivatives, as discussed by \cite{2023ApJ...958...84K, 2025PASJ...77.1323K}. Initial mass ratios were determined by averaging the estimates from both methods, which were found to be in close agreement (Table \ref{tab-solutions}). The $q$-search diagrams and the Kouzuma method process are presented in Figure \ref{Fig:q}. These values were subsequently adopted as the initial estimates for the light curve solution procedure, and then the final system mass ratios were determined during the subsequent modeling iterations.

The light curve of V1284 Her exhibits an asymmetry between its two maxima, commonly referred to as the O'Connell effect. Accurate modeling of this system required adding a cool starspot on the primary component (Table \ref{tab-solutions}). Such an effect is generally linked to magnetic activity on the stellar surface and is explained by the presence of starspots, although other physical mechanisms have also been proposed in previous studies (e.g., \citealt{1990ApJ...355..271Z, 2003ChJAA...3..142L}).
Given the degeneracy of spot solutions based solely on photometric data, the light curve analysis of V1284 Her was also carried out without including any starspot. The parameters derived from this solution show close agreement with those obtained from the spotted model, indicating that the inclusion of a spot does not significantly affect the fundamental system parameters (Table \ref{tab-solutions}). This consistency suggests that the observed O’Connell effect, although clearly detectable in all filters, does not impose strong constraints on the physical properties of the system within the quality of the present data. It should be noted that different starspot configurations in contact binary systems, such as their location, size, temperature contrast, and number, can produce similar photometric signatures. Ultimately, the spotted model of V1284 Her is adopted as the preferred solution to reproduce the O’Connell effect, while the non-spotted model is used as an independent consistency check of the results. In addition, we tested an alternative configuration in which a hot spot was placed on the secondary component; however, this setup did not yield an acceptable representation of the observed light curve and resulted in a noticeably poorer match to the data. It should be emphasized that these spot parameters represent only one plausible modeling scenario adopted to reproduce the O’Connell effect, and they are not uniquely constrained by the available photometric data.

\begin{figure*}
\centering
\includegraphics[width=0.99\textwidth]{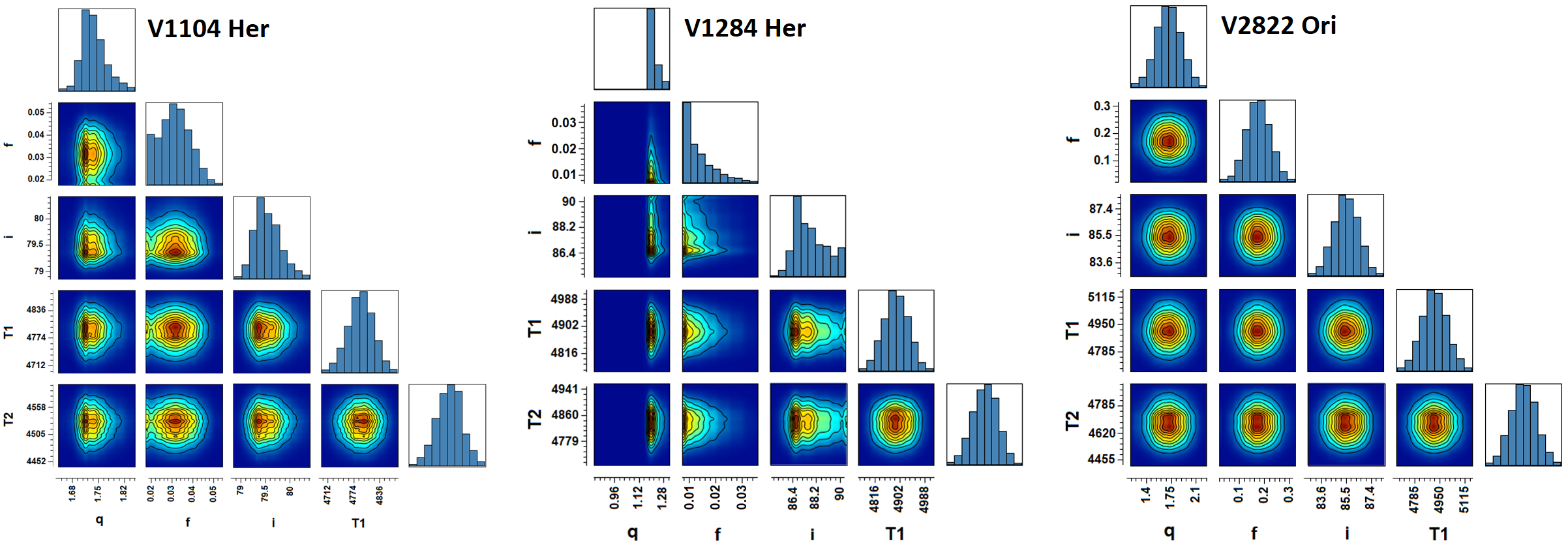}
\caption{Corner plots showing the posterior probability distributions and parameter correlations obtained from the MCMC analysis performed with the BSN application. The V1284 Her system was modeled with a cool starspot.}
\label{Fig:corner}
\end{figure*}

Using ground- and space-based photometric observations together with the initial parameter estimates, a theoretical fit to the light curves was constructed. Subsequently, the optimization module of the BSN application was employed to refine the solutions and to obtain tighter constraints on the component effective temperatures, mass ratio, fillout factor ($f$), and orbital inclination ($i$). Final parameter values and their associated uncertainties were derived through the MCMC analysis. Efficient MCMC performance is provided by the BSN application, enabling rapid generation of synthetic light curves during the fitting process. Twenty-four walkers were adopted, and 2500 iterations were executed to sample the five primary parameters ($T_1$, $T_2$, $q$, $f$, and $i$). The initial 400 iterations of each walker were excluded as burn-in to ensure proper convergence of the chains. The resulting posterior distributions obtained from the remaining samples were then used to determine the best-fitting parameter values and their corresponding 1$\sigma$ uncertainties, yielding a robust assessment of the model reliability. It should be noted that the quoted uncertainties of the temperatures and other main parameters derived from the light curve solution correspond to the formal errors obtained from the photometric modeling; the actual physical uncertainties may be larger due to the absence of spectroscopic constraints. Figure \ref{Fig:corner} presents the corner plots derived from the MCMC analysis, highlighting the posterior distributions and correlations among the model parameters. 

Light curve modelling of V2822 Ori included an additional third-light parameter $l_3$ to account for flux contamination from nearby sources within the photometric aperture, particularly in the TESS observations. A nearby field star (HD 291246) with a magnitude of approximately 10.58 contributes noticeable background light to the aperture. In addition, a fainter star (RA. $92.065178^\circ$, Dec. $-1.532200^\circ$) located very close to the target system in the sky background may also contribute to the observed flux contamination. Since these nearby sources are not physically associated with the eclipsing binary and do not introduce intrinsic photometric variability, their effect was treated as an additional constant light contribution through the $l_3$ parameter. To estimate the most appropriate third-light contribution, a dedicated $l_3$ search was performed over the range from 0.001 to 0.3. The optimal value was selected based on obtaining a better agreement between the synthetic and observed light curves together with a lower $\chi^2$ value. Although these contaminating sources may affect both the TESS and ground-based observations, their influence is expected to be stronger on the TESS pixels than on the ground-based $r'$-band data. The synthetic light curves obtained both with and without the inclusion of $l_3$ reproduce the observational data reasonably well; however, the MCMC corner plots indicate a better-constrained and more reliable solution for the model including the third-light contribution in both filters. An example image showing the field of view of V2822 Ori from our photometric observations is provided in the supplementary material of this study.

\begin{figure*}
\centering
\includegraphics[width=0.99\textwidth]{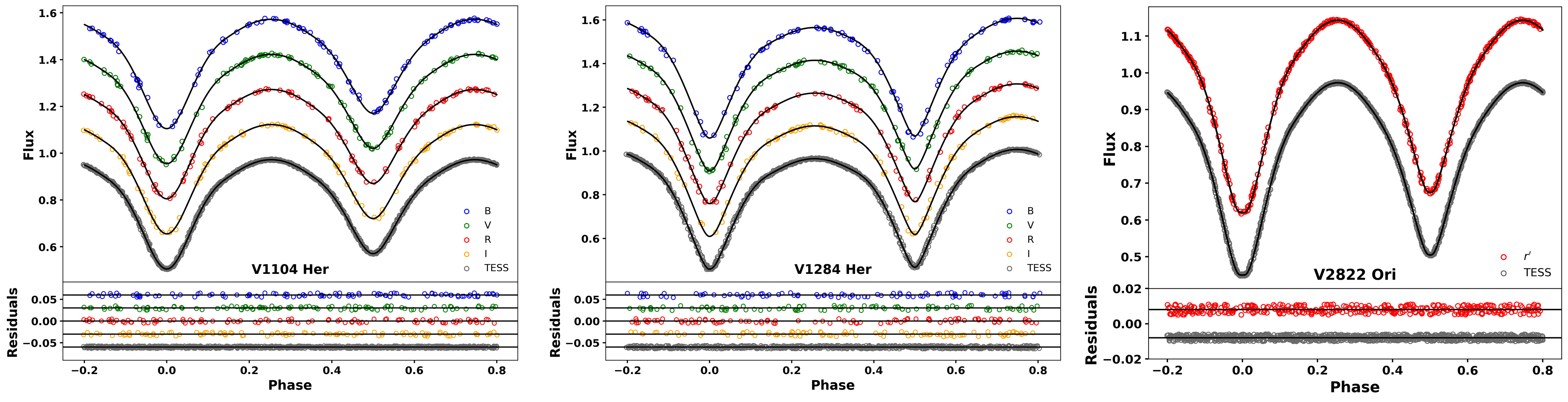}
\caption{The observed light curves of the three systems, obtained in different photometric bands, are shown with colored markers, together with the synthetic light curves computed from the best-fitting solutions. The corresponding residuals are presented at the lower part of each panel.}
\label{Fig:lc}
\end{figure*}

\begin{figure*}
\centering
\includegraphics[width=0.99\textwidth]{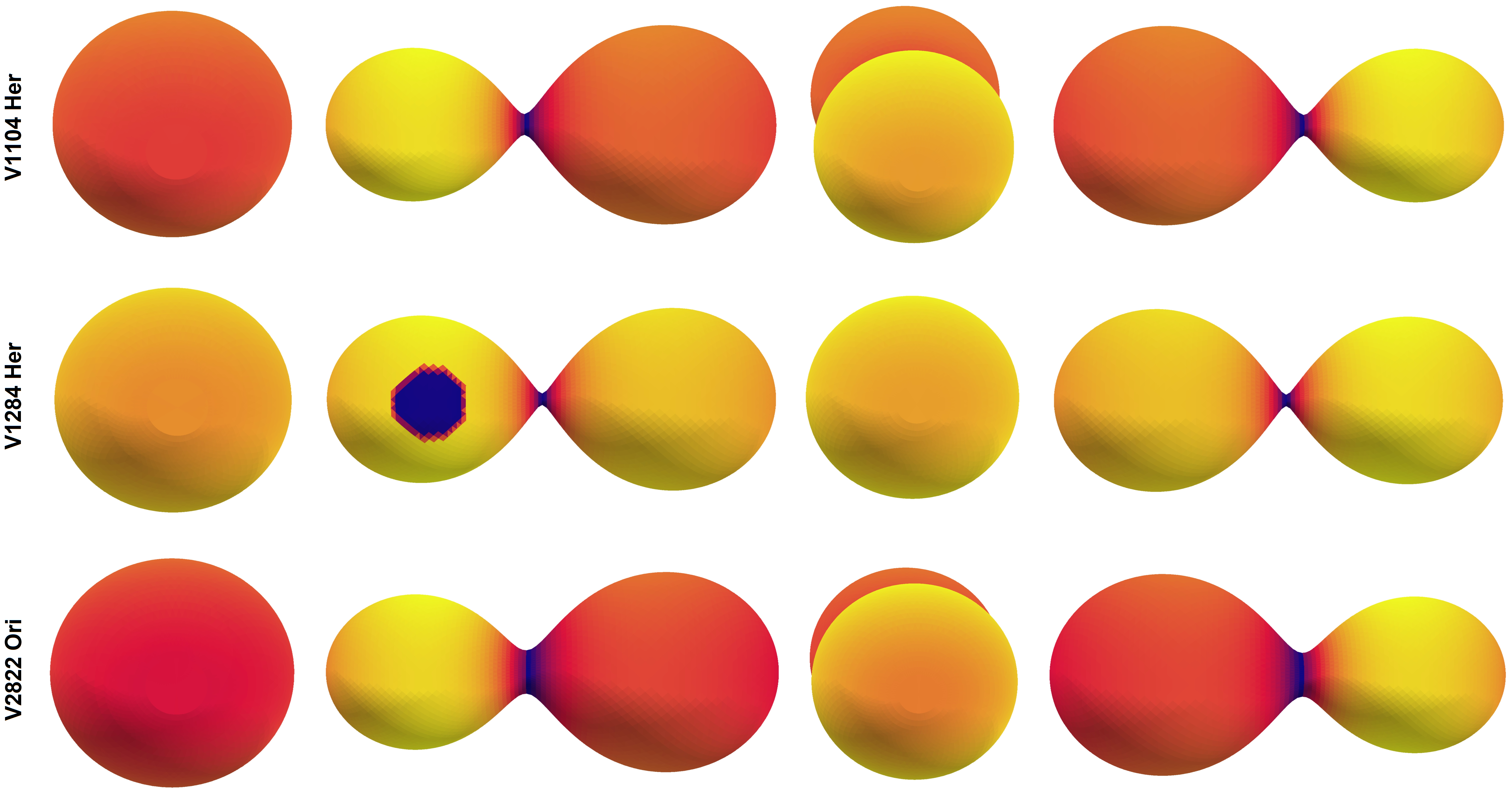}
\caption{Three-dimensional representations of the components of the target binary systems at four orbital phases. From left to right, the orbital phases are 0.00, 0.25, 0.50, and 0.75.}
\label{Fig:3d}
\end{figure*}

Following the determination of the main parameters by MCMC, others such as mean fractional radii $r_{\mathrm{mean}_{1,2}}$, are calculated internally by the BSN application. The mean fractional radii quantify the ratio of each star’s physical radius to the orbital semi-major axis of the binary star, offering a dimensionless measure of their size relative to the system’s overall geometry. This parameter is derived from the photometric light curve, where the depths and durations of eclipses, as well as the shape of the out-of-eclipse variations, directly depend on the stellar radii relative to the orbital separation. A summary of the light curve modeling results, including the inferred parameters and their uncertainties, is given in Table \ref{tab-solutions}. The final synthetic light curves, superimposed on the observed photometric measurements for all target binaries, are displayed in Figure \ref{Fig:lc}. Three-dimensional (3D) representations of the modeled binary systems are also shown in Figure \ref{Fig:3d}, with brighter surface colors denoting hotter regions and darker tones representing cooler areas.

All target systems are classified as contact binaries in various catalogs; although some catalogs, such as ASAS classification catalog (\citealt{2012ApJS..203...32R}), rely on the morphology of the light curves for classification. As listed in Table \ref{tab-solutions}, the resulting fillout factor for V1284 Her is extremely low, $f = 0.008^{+0.005}_{-0.002}$, indicating only a marginal degree of contact. To further investigate the nature of the system, an alternative light curve solution assuming a semi-detached configuration was also explored (\citealt{2023RAA....23e5005T}). Notably, the semi-detached model provided a visually acceptable fit, and the derived physical parameters showed no significant differences compared to those obtained under the contact assumption.

To distinguish between the two configurations in a quantitative and objective manner, the $\chi^2$ statistic was adopted as the primary criterion for model comparison. The $\chi^2$ value is defined as
\begin{equation}\label{chi2}
\chi^2 = \sum_{i=1}^{N} \frac{\left(F_{\mathrm{obs},i} - F_{\mathrm{model},i}\right)^2}{\sigma_i^2},
\end{equation}
where $F_{\mathrm{obs},i}$ represents the observed flux at the $i$-th data point, $F_{\mathrm{model},i}$ denotes the corresponding synthetic flux computed from the model, $\sigma_i$ is the observational uncertainty associated with each data point, and $N$ is the total number of observations. This statistic provides a measure of the overall discrepancy between the observed and modeled light curves, with lower values indicating a better fit to the data.

The reduced $\chi^2$ value obtained for the semi-detached configuration is 0.40, while a substantially lower value of 0.12 is found for the contact configuration, demonstrating a significantly improved agreement with the observational data. In addition to the numerical comparison, the residuals of the contact solution exhibit no systematic trends or cyclic structures, suggesting that the model adequately reproduces the essential features of the observed light curves. Conversely, the residuals corresponding to the semi-detached configuration display noticeable sinusoidal patterns, indicative of unmodeled systematics and a comparatively weaker representation of the data. These considerations strongly favor the contact configuration as the more reliable solution for V1284 Her.

\begin{table*}
\renewcommand\arraystretch{1.4}
\caption{Photometric parameters inferred from the light-curve analysis of the target systems.}
\centering
\begin{center}
\footnotesize
\begin{tabular}{c| c| c c| c}
 \hline
Parameter & V1104 Her & \multicolumn{2}{c|}{V1284 Her}  & V2822 Ori\\
& & With Starspot & Without Starspot & \\
\hline
$T_{1}$ (K) & $4392_{\rm-(32)}^{+(30)}$ & $4888_{\rm-(43)}^{+(44)}$ & 4893(48) & $4914_{\rm-(82)}^{+(83)}$\\
$T_{2}$ (K) & $4129_{\rm-(29)}^{+(24)}$ & $4830_{\rm-(42)}^{+(39)}$ & 4836(45) & $4679_{\rm-(84)}^{+(81)}$\\
$q=M_2/M_1$ & $1.736_{\rm-(31)}^{+(38)}$ & $1.201_{\rm-(15)}^{+(39)}$ & 1.295(46) & $1.709_{\rm-(175)}^{+(180)}$\\
$i^{\circ}$ & $79.52_{\rm-(22)}^{+(30)}$ & $86.40_{\rm-(92)}^{+(1.56)}$ & 86.49(1.21) & $85.49_{\rm-(96)}^{+(97)}$\\
$f$ & $0.033_{\rm-(8)}^{+(7)}$ & $0.008_{\rm-(2)}^{+(5)}$ & 0.013(5) & $0.170_{\rm-(46)}^{+(48)}$\\
$\Omega_1=\Omega_2$ & 4.854(89) & 4.058(95) & 4.208(88) & 4.734(78)\\
$l_1/l_{\mathrm{tot*}}$ & 0.431(7) & 0.462(6) & 0.453(6) & 0.431(4)\\
$l_2/l_{\mathrm{tot}}$ & 0.569(7) & 0.538(7) & 0.547(7) & 0.559(4)\\
$l_3/l_{\mathrm{tot}}$ & & & & 0.01(1)\\
$r_{1(mean)}$ & 0.333(5) & 0.364(6) & 0.358(6) & 0.347(7)\\
$r_{2(mean)}$ & 0.430(6) & 0.396(5) & 0.404(5) & 0.441(7)\\

$Col._{\text{spot}}$(deg) &  & 90(1) &  & \\
$Long._{\text{spot}}$(deg) &  & 281(2) &  & \\
$Rad._{\text{spot}}$(deg) &  & 21(1) &  & \\
$T_{\text{spot}}/T_{\text{star}}$ &  & 0.80(1) &  & \\
${\mathrm Component}_{\mathrm{spot}}$ &  & Primary &  & \\
\hline
$q$-search Method** & 1.741/0.574 & 1.119/0.894 & & 1.709/0.585\\
$q$ Kouzuma Method*** & 0.589(45) & 0.869(49) & & 0.577(43)\\
\hline
\end{tabular}
\end{center}
\label{tab-solutions}
\footnotesize \textit{*} $l_{\mathrm{tot}}$ for V1104 Her and V1284 Her is $l_1+l_2$, and for V2822 Ori is $l_1+l_2+l_3$.\\
\footnotesize \textit{**} Both $q$ and $1/q$ values are presented, separated by a slash (/), for better comparison.\\
\footnotesize \textit{***} In the Kouzuma method, the more massive component is consistently designated as $M_1$, and the less massive component as $M_2$.
\end{table*}

\vspace{0.6cm}
\section{Estimation of Absolute Parameters}
Absolute parameters for the three investigated systems were inferred using an empirical relation between the orbital period and the semi-major axis. An updated form of the $P$–$a$ relation was introduced by \cite{2024PASP..136b4201P}, based on a sample of 414 contact binaries with orbital periods shorter than 0.7 days (Equation \ref{eqPa}):

\begin{equation}\label{eqPa}
a=(0.372_{\rm-0.114}^{+0.113})+(5.914_{\rm-0.298}^{+0.272})\times P.
\end{equation}

Derivation of the semi-major axis $a$ (in units of $R_\odot$) was performed using the orbital period $P$ expressed in days. Stellar masses were subsequently constrained by combining the calculated semi-major axis with the orbital period and the photometrically determined mass ratio. The corresponding masses were then derived from Kepler's Third Law, as given by Equations \ref{eq:M1} and \ref{eq:M2}.

\begin{equation}\label{eq:M1}
M{_1}=\frac{4\pi^2a^3}{GP^2(1+q)},
\end{equation}

\begin{equation}\label{eq:M2}
M{_2}=q\times{M{_1}}.
\end{equation}

Estimates of the stellar radii were obtained by scaling the fractional mean radii $r_{\mathrm{mean}1,2}$ derived from the light curve modeling with the semi-major axis, as described by Equation \ref{eqR}:

\begin{equation}\label{eqR}
R_{1,2}=a\times r_{(1,2mean)}.
\end{equation}

Luminosities of both components were then evaluated using their radii and effective temperatures through the standard stellar relation (Equation \ref{eqL}):

\begin{equation}\label{eqL}
L_{1,2}=4\pi R^2 \sigma T_{1,2}^4.
\end{equation}

Using the derived luminosities, the absolute bolometric magnitudes ($M_{\mathrm{bol}}$) were calculated based on the well-known relation between luminosity and bolometric magnitude (Equation \ref{eqMbol}). A solar bolometric magnitude of $4.73\, \mathrm{mag.}$ was adopted following \cite{2010AJ....140.1158T}.

\begin{equation}\label{eqMbol}
M_{{\rm bol},1,2}=M_{{\rm bol},\odot}-2.5\log\left(\frac{L_{1,2}}{L_\odot}\right).
\end{equation}

Finally, the logarithmic surface gravity ($\log g$) was calculated from the estimated stellar masses and radii using the solar-scaled relation (Equation \ref{eqLogg}), where $\log g_\odot = 4.438$ (cgs), and $M$ and $R$ are expressed in units of $M_\odot$ and $R_\odot$, respectively.

\begin{equation}\label{eqLogg}
\log g=\log g_\odot+\log\left(\frac{M}{M_\odot}\right)
-2\log\left(\frac{R}{R_\odot}\right).
\end{equation}

Moreover, the orbital angular momentum ($J_0$) of each system was evaluated using Equation \ref{eqJ0} from \cite{2006MNRAS.373.1483E}, where $q$ denotes the mass ratio, $M$ is the total system mass, $P$ is the orbital period, and $G$ is the gravitational constant.

\begin{equation}\label{eqJ0}
J_0=\frac{q}{(1+q)^2} \sqrt[3]{\frac{G^2}{2\pi}M^5P}.
\end{equation}

The uncertainties of the results were calculated considering the errors of the equation \ref{eqPa} and then associated parameters in the computation process. Finally, the absolute parameters estimated for the target binary systems are presented in Table \ref{tab:absolute}.

\begin{table*}
\renewcommand\arraystretch{1.2}
\caption{Absolute parameter estimation for the target systems.}
\centering
\begin{center}
\begin{tabular}{c c c c c}
\hline
Parameter & V1104 Her & V1284 Her & V2822 Ori\\
\hline
$M_1 (M_\odot)$ & 0.48(17) & 0.71(21) & 0.50(17)\\
$M_2 (M_\odot)$ & 0.83(31) & 0.85(27) & 0.86(40)\\
$R_1 (R_\odot)$ & 0.57(7) & 0.86(9) & 0.64(8)\\
$R_2 (R_\odot)$ & 0.74(9) & 0.94(10) & 0.81(10)\\
$L_1 (L_\odot)$ & 0.11(3) & 0.38(10) & 0.21(7)\\
$L_2 (L_\odot)$ & 0.14(4) & 0.43(11) & 0.28(10)\\
$M_{\mathrm{bol1}}$(mag.) & 7.13(28) & 5.78(26) & 6.40(32)\\
$M_{\mathrm{bol2}}$(mag.) & 6.84(27) & 5.65(25) & 6.10(32)\\
$\log g_1$(cgs) & 4.60(3) & 4.42(2) & 4.53(2)\\
$\log g_2$(cgs) & 4.62(4) & 4.43(4) & 4.55(7)\\
$a(R_\odot)$ & 1.72(18) & 2.37(21) & 1.84(18)\\
$logJ_0$(cgs) & 51.44(22) & 51.65(19) & 51.48(24)\\
\hline
\end{tabular}
\end{center}
\label{tab:absolute}
\end{table*}

For comparison, we also calculated the absolute parameters using the Gaia DR3 parallax method comprehensively described by \cite{2024NewA..11002227P}. This approach provides an independent estimate of the stellar parameters based on geometric distances. The extinction values ($A_V$) for the investigated systems were estimated using the three-dimensional dust maps of \citet{2019ApJ...887...93G} together with Gaia DR3 distance estimates, yielding $A_V = 0.074(1)$ for V1104 Her, $0.048(1)$ for V1284 Her, and $0.180(1)$ for V2822 Ori in magnitude unit. Since the derived $A_V$ values are low for all three binaries, the Gaia DR3 parallax method can be reliably applied (\citealt{2024NewA..11002227P}). The absolute visual magnitude of each system was determined from the Gaia DR3 distance, the observed $V_{\rm max}$ (Table \ref{tab:systemsinfo}), and the corresponding $A_V$ value. Then the individual component visual magnitudes were derived using the luminosity ratios obtained from the multiband light curve solutions. After applying the bolometric corrections from \cite{1996ApJ...469..355F}, the bolometric magnitudes and stellar luminosities were calculated using well-known astrophysical equations. The stellar radii were subsequently determined from the derived luminosities and effective temperatures. The semi-major axis was then calculated using the stellar radii together with the corresponding mean fractional radii obtained from the light-curve solutions, and the component masses were finally derived from Kepler's third law and mass ratio. The Gaia DR3 parallax method yields semi-major axes of $1.748(92)\,R_{\odot}$, $2.242(114)\,R_{\odot}$, and $1.527(141)\,R_{\odot}$ for V1104 Her, V1284 Her, and V2822 Ori, respectively. The corresponding component masses are $M_1 = 0.504(80)\,M_{\odot}$ and $M_2 = 0.876(139)\,M_{\odot}$ for V1104 Her, $M_1 = 0.604(92)\,M_{\odot}$ and $M_2 = 0.726(111)\,M_{\odot}$ for V1284 Her, and $M_1 = 0.286(81)\,M_{\odot}$ and $M_2 = 0.490(138)\,M_{\odot}$ for V2822 Ori. The independently derived parameters are in excellent agreement with those obtained from the empirical $P-a$ relation for V1104 Her and V1284 Her, confirming our results. For V2822 Ori, the Gaia DR3 parallax results remain reasonably consistent with the primary solution, although a somewhat larger difference is found. This discrepancy is most likely attributable to the third-light contribution detected in the light-curve analysis, which affects the luminosity determination and consequently propagates into the Gaia DR3 parallax solution.

\vspace{0.6cm}
\section{Investigation of Orbital Period Variations}
The investigation of orbital period variations for the targets was carried out using the observed minus calculated (O-C) eclipse-time method. Eclipse timing measurements were assembled from several photometric surveys, namely VSX, TESS, and VarAstro\footnote{\url{https://var.astro.cz/en/}}. Eclipse timings were determined directly from the TESS 2-min, and TESS 10-min cadence data. For the TESS 30-min cadence observations, the correction procedure by \cite{2020AJ....159..189L} was applied by shifting the data by one orbital cycle prior to the determination of the eclipsing times. To extract the times of minima for the target systems, an analytical model based on Gaussian distribution was fitted to selected regions of the light curves containing the eclipses. Methods such as the Kwee-van Woerden (KW; \citealt{1956Kwee}) technique may become unreliable when the observed minima are asymmetric or when the light curves are incomplete, as discussed by \cite{2015A&AMikul}. Parameter estimation and uncertainty calculations were performed using MCMC sampling methods, while the reported uncertainties correspond to the $1\sigma$ confidence intervals derived from the posterior distributions. Conversion of all Heliocentric Julian Dates (HJD) to Barycentric Julian Dates (BJD) in the Barycentric Dynamical Time system was performed using the online tool developed by \cite{2010PASP..122..935E}. The mean uncertainty derived from the used ground-based minima was adopted for CCD-based minima reported by VarAstro and VSX in cases where individual error estimates were not provided. A summary of eclipse timings extracted from our observations, the literature, and a subset of TESS data derived from our photometric analysis is presented in Table \ref{Tab:extracted-mins}. An extensive machine-readable archive containing the complete set of times of minima for the analyzed binary systems is made available online.

\begin{table*}
\caption{The measured eclipse timings for the three studied binaries. The continuation of the extracted eclipse timings from TESS is available in the online version.}
\centering
\small
\begin{tabular}{c c c c| c c c c}
\hline
Min.($BJD_{TDB}$) & Epoch & O-C & Reference & Min.($BJD_{TDB}$) & Epoch & O-C & Reference\\ 
\hline
V1104 Her	&		&		&		&	V1284 Her	&		&		&		\\
2452526.4294	&	-36500	&	-0.0036	&	VSX	&	2452898.3483	&	-23561	&	0.0229	&	VSX	\\
2451275.8432(4)	&	-41988	&	-0.0065	&	1	&	2451308.8759(8)	&	-28275.5	&	0.0253	&	3	\\
2451358.1076(3)	&	-41627	&	-0.0054	&	12	&	2451311.7418(9)	&	-28267	&	0.0255	&	3	\\
2452548.3059(12)	&	-36404	&	-0.0032	&	2	&	2452871.3760(7)	&	-23641	&	0.0222	&	3	\\
2452840.4468(5)	&	-35122	&	0.0008	&	4	&	2453107.5438(20)	&	-22940.5	&	0.0192	&	4	\\
2453620.4618	&	-31699	&	-0.0035	&	VarAstro	&	2453117.4968(30)	&	-22911	&	0.0264	&	4	\\
2454000.3340(6)	&	-30032	&	-0.0005	&	5	&	2453203.4638	&	-22656	&	0.0212	&	VarAstro	\\
2454971.4270(8)	&	-25770.5	&	-0.001	&	6	&	2453410.6387	&	-22041.5	&	0.02	&	VarAstro	\\
2455067.3643(8)	&	-25349.5	&	0.0006	&	7	&	2453988.3358(3)	&	-20328	&	0.0173	&	5	\\
2455672.3735(9)	&	-22694.5	&	-0.0009	&	9	&	2454217.4266(4)	&	-19648.5	&	0.0174	&	5	\\
2455741.7623(4)	&	-22390	&	-0.0003	&	8	&	2457513.8731(1)	&	-9871	&	0.0189	&	14	\\
2457157.7802(2)	&	-16176	&	-0.0035	&	13	&	2457515.3917(31)	&	-9866.5	&	0.0204	&	15	\\
2457518.5083(3)	&	-14593	&	-0.0031	&	15	&	2457515.5589(5)	&	-9866	&	0.019	&	15	\\
2457629.4839(1)	&	-14106	&	0.0031	&	16	&	2458983.8216(18)	&	-5511	&	0.0109	&	TESS	\\
2458984.6667(2)	&	-8159	&	0.0015	&	TESS	&	2458983.9828(16)	&	-5510.5	&	0.0035	&	TESS	\\
2460843.6714(25)	&	-1	&	0.0012	&	This study	&	2458984.1591(3)	&	-5510	&	0.0112	&	TESS	\\
2460843.7817(30)	&	-0.5	&	-0.0015	&	This study	&	2460841.8223(9)	&	0	&	0	&	This study	\\
2460843.9051(19)	&	0	&	0	&	This study	&	2460841.9904(33)	&	0.5	&	-0.0005	&	This study	\\
V2822 Ori	&		&		&		&		&		&		&		\\
2455523.9334(11)	&	-20633.5	&	-0.0002	&	8	&	2460642.6386(22)	&	0	&	0	&	This study	\\
2455958.6879(3)	&	-18881	&	-0.0013	&	10	&	2460642.7641(5)	&	0.5	&	0.0015	&	This study	\\
2456256.8772(22)	&	-17679	&	-0.0011	&	11	&	2460644.7505(4)	&	8.5	&	0.0032	&	This study	\\
2458442.9340	&	-8867	&	-0.0023	&	VSX	&	2460686.6716(7)	&	177.5	&	-0.0008	&	TESS	\\
2458821.0060	&	-7343	&	-0.0003	&	VarAstro	&	2460686.7959(9)	&	178	&	-0.0005	&	TESS	\\
2459202.0551(14)	&	-5807	&	0.0019	&	TESS	&	2460687.0458(8)	&	179	&	0.0014	&	TESS	\\
2459202.1788(9)	&	-5806.5	&	0.0016	&	TESS	&	2460687.1708(12)	&	179.5	&	0.0023	&	TESS	\\
\hline
\end{tabular}
\label{Tab:extracted-mins}\\
\footnotesize References: 1=\cite{2002IBVS.5333....1B}, 2=\cite{2003IBVS.5438....1D}, 3=\cite{2004IBVS.5516....1B}, 4=\cite{2004IBVS.5543....1D}, 5=\cite{2007IBVS.5781....1D}, 6=\cite{2010IBVS.5918....1H}, 7=\cite{diethelm2010timings}, 8=\cite{diethelm2011timings}, 9=\cite{2012IBVS.6010....1H}, 10=\cite{diethelm2012timings}, 11=\cite{diethelm2013timings}, 12=\cite{2015AJ....149..148L}, 13=\cite{2016IBVS.6164....1N}, 14=\cite{2017IBVS.6195....1N}, 15=\cite{2017IBVS.6196....1H}, 16=\cite{loukaidou2022cobitom}\\
\end{table*}

The O-C values were evaluated using the following relation
\begin{equation}
\mathrm{BJD} = \mathrm{BJD}_{0} + P \times E,
\end{equation}
in which $\mathrm{BJD}$ denotes the observed eclipse timing, $\mathrm{BJD}_{0}$ represents the reference epoch, $P$ is the orbital period (both parameters are given in Table \ref{tab:ephemeris}), and $E$ corresponds to the cycle number. Derived O-C values are presented in Table \ref{Tab:extracted-mins}, together with a machine-readable version of the dataset. A new ephemeris was calculated for each system, and the corresponding parameters are provided in Table \ref{tab:ephemeris}. The linear ephemeris was represented by the relation
\begin{equation}
\mathrm{O\!-\!C} = \Delta T_{0} + \Delta P_{0} \times E.
\end{equation}

In addition to the linear ephemeris, a sinusoidal model was examined to test for the presence of possible cyclic period variations. Such variations may arise from several physical mechanisms, including the light-travel time effect induced by an additional body in the system or magnetic activity cycles operating through the Applegate mechanism (\citealt{1992ApJ...385..621A}). The adopted sinusoidal model provides a phenomenological description of the observed modulation and, by itself, does not distinguish between these possible physical origins. The adopted model has the form
\begin{equation}
\mathrm{O\!-\!C}=\Delta T_{0}+\Delta P_{0}E+
A\sin\left(\frac{2\pi E}{P_{\rm mod}}+\phi\right),
\end{equation}
where $A$ is the semi-amplitude of the modulation, $P_{\rm mod}$ is the modulation period expressed in orbital cycles, and $\phi$ is the phase. Consequently, the linear model contains two free parameters ($\Delta T_{0}$ and $\Delta P_{0}$), whereas the cyclic model contains five free parameters ($\Delta T_{0}$, $\Delta P_{0}$, $A$, $P_{\rm mod}$, and $\phi$).

Both models were fitted to the complete eclipse-timing datasets using weighted least squares. Individual timing uncertainties were adopted as weights whenever available. For eclipse timings without published uncertainties, the mean uncertainty derived from the available ground-based CCD minima was assigned. The goodness of fit was evaluated using the same $\chi^2$ statistic defined in Equation~\ref{chi2}. In this case, $F_{\mathrm{obs},i}$ and $F_{\mathrm{model},i}$ were replaced by the observed and model-predicted $O\!-\!C$ values, respectively, while $N$ denotes the number of eclipse timings and $\sigma_i$ represents the uncertainty associated with each timing measurement.

To compare the competing models while accounting for the different numbers of free parameters, the Akaike Information Criterion (AIC) and the Bayesian Information Criterion (BIC) were calculated according to
\begin{equation}
{\rm AIC}=\chi^{2}+2k,
\end{equation}

\begin{equation}
{\rm BIC}=\chi^{2}+k\ln N,
\end{equation}
where $k$ is the number of free parameters and $N$ is the number of eclipse timings used in the fit. The numbers of eclipse timings are $N=452$, $N=399$, and $N=130$ for V1104 Her, V1284 Her, and V2822 Ori, respectively. The resulting AIC, BIC, and $\chi^2$ values are summarized in Table~\ref{tab:aic_bic}.

\begin{figure*}
\centering
\includegraphics[width=0.99\textwidth]{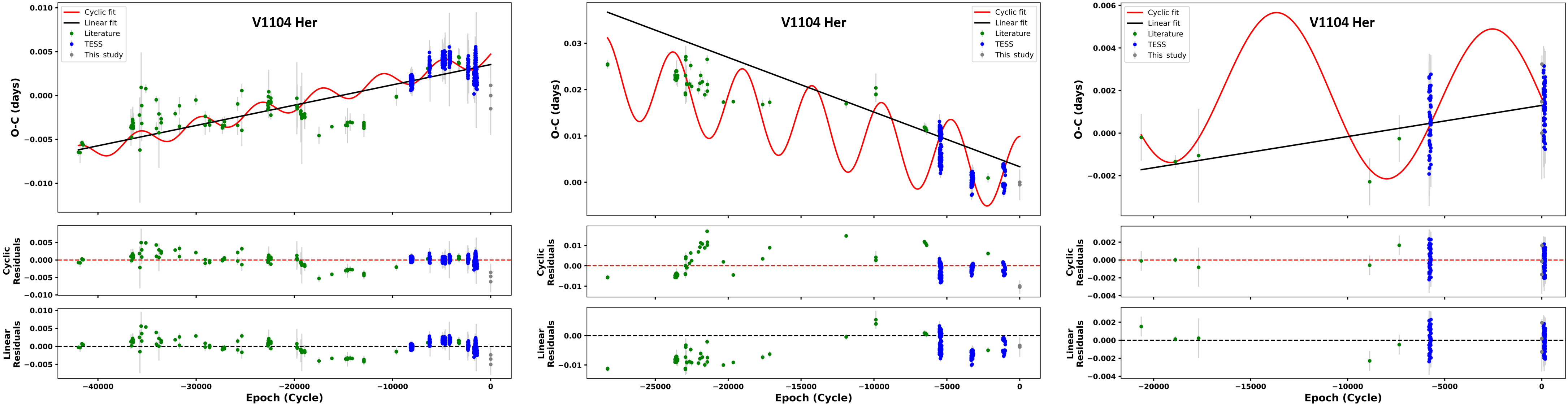}
\caption{O-C diagrams for the target systems based on linear fits. The lower panel shows the residuals.}
\label{Fig:oc}
\end{figure*}

The O-C diagram corresponding to the targets is displayed in Figure \ref{Fig:oc}.

\renewcommand\arraystretch{1.2}
\begin{table*}
\centering
\small
\caption{Reference and new ephemerides for the targets. The orbital period $P_{0}$ comes from VSX.}
\begin{tabular}{c c c c c}
\hline
System & $t_0$($BJD_{TDB}$) & $P$(d) & Corrected $t_0$($BJD_{TDB}$) & Corrected $P$(d)\\
\hline
V1104 Her & 2460843.9051(20) & 0.22787595 & 2460843.9095(2) & 0.22787621(3) \\
V1284 Her & 2460841.8223(11) & 0.33714600 & 2460841.8238(2) & 0.33714524(8) \\
V2822 Ori & 2460642.6386(32) & 0.24807740 & 2460642.6399(2) & 0.24807755(2) \\
\hline
\label{tab:ephemeris}
\end{tabular}
\end{table*}

The apparent morphology of the O-C diagrams suggested that both linear and cyclic models should be examined for each target. Previous studies have also suggested possible cyclic behavior for V1104 Her (\citealt{2015AJ....149..148L}; \citealt{loukaidou2022cobitom}), providing additional motivation for testing the sinusoidal model.

Based on the statistical comparison summarized in Table~\ref{tab:aic_bic}, the cyclic model provides a substantially better fit than the linear ephemeris for V1104 Her and V1284 Her, as evidenced by the significantly lower AIC, BIC, and $\chi^2$ values. These results indicate that a periodic modulation offers a more adequate description of the currently available eclipse timings for these two systems. Nevertheless, as mentioned above, a sinusoidal modulation is not unique evidence for the presence of a tertiary companion, since other mechanisms, such as magnetic activity cycles, may also produce cyclic O-C variations. For V2822 Ori, the statistical comparison is less decisive. Although the cyclic model yields a slightly lower AIC value, the linear model is preferred according to the BIC owing to its smaller number of free parameters. Since the differences between the two models are relatively small, the sinusoidal solution was also retained to allow a consistent comparison with the other systems.

Assuming that the observed cyclic variations are produced by the light-travel time effect, the orbital parameters of the proposed tertiary companions derived from the best-fitting sinusoidal solutions are presented in Table~\ref{tab:sin3}. For V2822 Ori, although the linear model is preferred, the relatively small difference between the linear and cyclic models (Table~\ref{tab:aic_bic}) motivated us to also report the parameters derived under the light-travel time effect interpretation in Table~\ref{tab:sin3}. More generally, the inferred modulation periods are long compared with the current observational baselines, and continued eclipse-timing observations will be required to determine the physical origin of the observed modulations and to confirm or refute the proposed tertiary companions.

\begin{table}
\centering
\caption{AIC, BIC and $\chi^2$ values for linear and cyclic fits of the three systems.}
\begin{tabular}{c c c c c c}
\hline
System & Fit type & AIC & BIC  & $\chi^2$ \\
\hline
V1104 Her & Linear & 1246.2 & 1254.4 & 7057.2\\
                    & Cyclic & 1169.1 & 1189.7 & 5872.8\\
V1284 Her & Linear & 2169.5 & 2177.5 & 90787.4\\
                   & Cyclic & 2108.3 & 2128.3 & 76719.1\\
V2822 Ori & Linear & 64.1 & 69.8 & 206.4\\
                  & Cyclic & 63.1 & 77.5 & 195.6\\
\hline
\label{tab:aic_bic}
\end{tabular}
\end{table}

\begin{table*}
\centering
\caption{Best-fitting sinusoidal parameters and the corresponding derived properties of the proposed third body for the investigated systems; all quoted uncertainties represent $1\sigma$ confidence intervals.}
\begin{tabular}{l c c c c}
\hline
Parameter & V1104 Her & V1284 Her & V2822 Ori & Unit\\
\hline
\multicolumn{5}{c}{\textit{Sinusoidal Fit Parameters}}\\
\hline
$a$ & $0.00441 \pm 0.00003$ & $0.00155 \pm 0.00006$ & $0.00101 \pm 0.00026$ & days\\
$b$ & $(2.64 \pm 0.03)\times10^{-7}$ & $(-7.64 \pm 0.08)\times10^{-7}$ & $(-6.9 \pm 7.1)\times10^{-8}$ & days\,cycle$^{-1}$\\
$A$ & $-0.00098 \pm 0.00003$ & $0.00841 \pm 0.00009$ & $0.00371 \pm 0.00122$ & days\\
$P_{\rm mod}$ & $6172.81 \pm 26.34$ & $4766.87 \pm 3.75$ & $11127.56 \pm 362.52$ & cycles\\
$\phi$ & $-2.829 \pm 0.053$ & $1.455 \pm 0.008$ & $-3.312 \pm 0.064$ & rad\\
\hline
\multicolumn{5}{c}{\textit{Derived Third-body Parameters}}\\
\hline
$a_{12}\sin i_3$ & $0.170 \pm 0.005$ & $1.455 \pm 0.016$ & $0.642 \pm 0.211$ & AU\\
$f(M_3)$ & $(1.7 \pm 0.2)\times10^{-5}$ & $0.0181 \pm 0.0006$ & $0.00029 \pm 0.00028$ &
$M_\odot$\\
$M_{3,\rm min}$ & $0.031$ & $0.414$ & $0.084$ & $M_\odot$\\
$M_3$ (MC mean / median / std) & $0.031 / 0.031 / 0.006$ & $0.412 / 0.413 / 0.056$ & $0.073 / 0.079 / 0.041$ & $M_\odot$\\
$M_3$ ($1\sigma$ interval) & $[0.026,\,0.037]$ & $[0.356,\,0.467]$ & $[0.021,\,0.112]$ & $M_\odot$\\
$P_{\rm mod}$ & $16.900 \pm 0.072$ & $13.051 \pm 0.010$ & $30.466 \pm 0.993$ & yr\\
\hline
\end{tabular}
\label{tab:sin3}
\end{table*}

\vspace{0.6cm}
\section{Discussion and Conclusion}
This investigation presents a photometric analysis of three W UMa-type contact binary systems based on a combination of ground- and space-based photometric data. We derived the light curve solutions of the three systems using the BSN application in combination with the MCMC method. In these analyses, we deliberately estimated the system parameters independently of the previous studies by \cite{2015AJ....149..148L} and \cite{loukaidou2022cobitom} for V1104 Her.

\begin{figure*}
\centering
\includegraphics[width=0.99\textwidth]{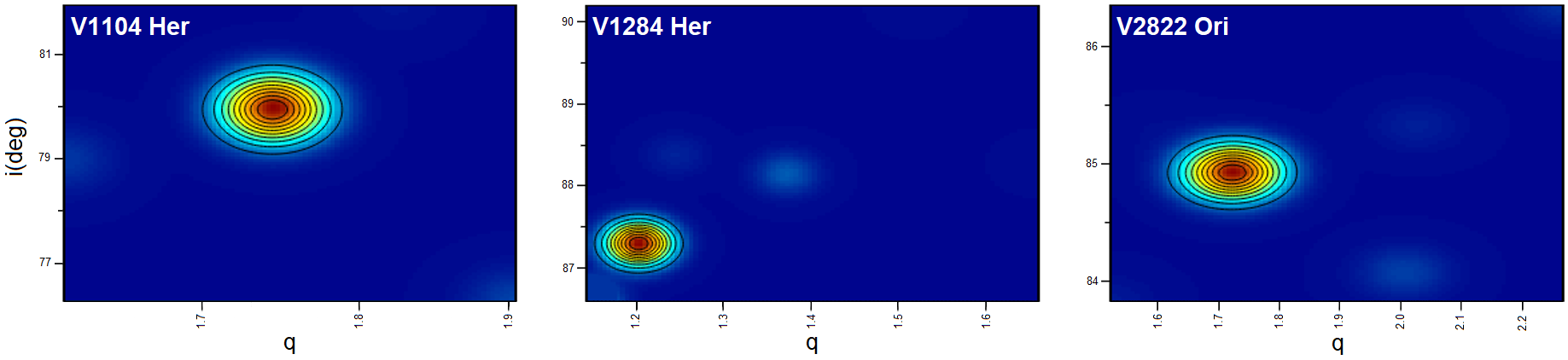}
\caption{Mass ratio and orbital inclination search using an MCMC approach to test a new capability of the BSN application. Results from this test are consistent with those obtained using the other methods applied in this study. The search was conducted over a range of $q=0.05-20$ and $i=30^{\circ}-90^{\circ}$; the displayed range has been subsequently narrowed to the high-probability (posterior) region identified by the BSN application.}
\label{Fig:q-test}
\end{figure*}

Initial mass ratios of the target systems were determined using several methods. A $q$-search procedure was performed first, followed by the application of the Kouzuma method, which yielded nearly consistent results. An additional $q$-search was carried out using an MCMC approach implemented in the BSN application as a test of its new capability, adopting 15 walkers and 500 iterations over a wide range (Figure \ref{Fig:q-test}); this approach also produced results consistent with the other methods. Final values of the mass ratio, along with their uncertainties, were derived after performing MCMC on all parameters. This consistency in mass ratio estimates across all photometric search methods underscores the reliability of our findings. The resulting mass ratios of the target systems indicate that the secondary components are more massive than the primaries. The results of the light curve solutions further show that all systems have high orbital inclinations. Evidence from earlier works, including \cite{2021AJLi} and \cite{2024AJPoro}, suggests that systems with high orbital inclinations and total eclipses allow for more reliable determination of the mass ratio.

A totally eclipsing binary system is generally defined as a system in which at least one eclipse is total, meaning that one stellar component is completely obscured by the other during eclipse. The eclipse morphology primarily depends on the orbital inclination and the Roche geometry of the system. A commonly used first-order approximation for the occurrence of total eclipses is given by $\cos i \lesssim |r_2-r_1|$ (\citealt{2020ApJS..247...50S}), where $i$ is the orbital inclination and $r_1$ and $r_2$ are the mean fractional radii of the components. However, this relation is derived under the assumption of spherical stars and therefore provides only an approximate geometrical guideline. In contact binaries, where both components fill their Roche lobes and are significantly distorted, the eclipse morphology is more reliably determined from the complete Roche geometry obtained from the light-curve solution. The derived inclinations and fractional radii of V1104 Her, V1284 Her, and V2822 Ori are listed in Table~\ref{tab-solutions}, while their corresponding three-dimensional Roche configurations are shown in Figure~\ref{Fig:3d}. Based on the approximate analytical criterion, V2822 Ori satisfies the condition for a total eclipse, whereas V1104 Her and V1284 Her lie close to the corresponding geometrical limit and are not clearly identified as totally eclipsing by this approximate relation alone. Nevertheless, the complete Roche-geometry solutions derived from the light curve modeling predict that all three systems undergo complete occultation of one stellar component at the primary minimum (phase 0.0), consistent with the corresponding three-dimensional Roche models shown in Figure~\ref{Fig:3d}. In addition, V1284 Her exhibits complete eclipses at both conjunctions. These eclipse configurations are illustrated by the corresponding three-dimensional Roche models and are consistent with the synthetic light curves. The flat-bottom morphology of the primary eclipse in V2822 Ori provides additional observational support for its total-eclipse configuration. Therefore, the eclipse classification adopted in this work is based on the complete Roche-geometry solutions rather than on the approximate analytical criterion alone, and all three systems are classified as totally eclipsing contact binaries.

Based on the classification scheme proposed by \cite{2022AJ....164..202L}, all systems are found to be the shallow-contact classification. Consequently, it can be concluded that, in particular, V1104 Her and V1284 Her exhibit very low degrees of contact. The light curve analysis shows that the effective temperatures of the stellar components in the selected systems range to below about 5000 K, indicating that the target systems consist of relatively cool stellar components. The temperature differences between the stellar components were found to be 263 K for V1104 Her, 58 K for V1284 Her, and 249 K for V2822 Ori, consistent with the general behavior observed in contact binary systems.

The initial effective temperatures adopted in this study for the target systems described in Section 3 were selected based on the availability and reliability of catalog values. For V1104 Her, the temperature provided by Gaia DR3 was used as the initial input for the modeling, as it is derived using updated methodologies that incorporate BP/RP spectrophotometry and more advanced parameter inference techniques (\citealt{de2023gaia}). The difference between this value and that from Gaia DR2 for V1104 Her is about 205 K, which lies within the typical uncertainties of such estimates (\citealt{andrae2018gaia}). For the other two systems, no temperature values are reported in Gaia DR3; therefore, values from TIC were adopted (Table \ref{tab:systemsinfo}). The differences between the TIC and Gaia DR2 temperatures for the V1284 Her and V2822 Ori systems are approximately 397 K and 219 K, respectively. These discrepancies are not unexpected, as the Gaia DR2 temperatures are primarily derived from broad-band photometry under simplified assumptions, particularly regarding interstellar extinction, whereas the TIC values are estimated using multiband photometric data and combined calibrations (\citealt{2019AJ....158..138S}). It should also be noted that all these catalog temperatures are derived under the assumption of single stars, while the systems under study are contact binaries. Therefore, systematic differences are expected due to the combined light of the two components. Consequently, these temperatures are used only as initial estimates to constrain the parameter space in the light curve modeling. Furthermore, we sought to derive an estimate for the systems' temperatures employing ground-based data in $B$ and $V$ filters. The interstellar reddening was determined using the 3D dust map of \cite{2019ApJ...887...93G}, adopting distances inferred from Gaia DR3 parallaxes (Table \ref{tab:systemsinfo}), yielding $E(B-V)=0.03$ and $0.05$ magnitude for V1104 Her and V1284 Her, respectively. The ($B-V$) color indices for V1104 Her and V1284 Her were calculated from our ground-based photometric observations. The intrinsic color indices were then obtained using $(B-V)_0=(B-V)-E(B-V)$. Therefore, $(B-V)_0$ for V1104 Her is 1.32, and V1284 Her is 1.01 magnitudes. Finally, the systems’ temperatures were estimated using the calibration tables of \cite{1996ApJ...469..355F} based on the derived intrinsic colors, resulting in temperatures of 4283 K, and 4825 K for V1104 Her, and V1284 Her, respectively. The estimated values are in good agreement with the temperatures in Table \ref{tab:systemsinfo}, which were used to initiate the light curve solution. Based on the effective temperature-spectral type relations reported by \cite{2018MNRAS.479.5491E}, the inferred spectral categories of the companion stars are K5–K5 for V1104 Her, K2–K2 for V1284 Her, and K2–K3 for V2822 Ori.

W UMa-type contact binaries are commonly classified into A- and W-subtypes, where the more massive component is hotter in A-subtype and cooler in W-subtype systems (\citealt{1970VA.....12..217B}). These subtypes are thought to follow different evolutionary pathways (\citealt{2020MNRAS.492.4112Z}). Our estimations show that, in all studied systems, the primary component is hotter but less massive than the secondary, indicating that all targets belong to the W-subtype. The results indicate that the studied stars have masses below $0.86\,M_\odot$, classifying them as low-mass stars (Table \ref{tab:absolute}).

\cite{2015AJ....149..148L} performed a photometric analysis of V1104 Her using multiband light curves modeled with the Wilson-Devinney code. Their results indicate that V1104 Her is a W-subtype shallow contact binary, with a degree of contact of about \( f \approx 15\% \). Therefore, the results of \cite{2015AJ....149..148L} are consistent with our results, indicating that V1104 Her is classified as the W-subtype. This study yields an inverse mass ratio for V1104 Her of $1/q = 0.576^{+0.012}_{-0.010}$, where the quoted uncertainties were obtained by propagating the asymmetric errors of the original mass ratio using a linear approximation. The result does not differ significantly from the mass ratio reported by \cite{2015AJ....149..148L}, particularly considering the significant time elapsed since their study, as well as the different observational datasets and analysis approaches employed. Although the identification of the more massive component differs between the two analyses, this discrepancy may be related not only to the evolutionary state and ongoing mass and energy transfer in contact systems, but also to differences in the modeling approach and the observational data used in the analyses.
\cite{loukaidou2022cobitom} also investigated V1104 Her within the framework of the CoBiToM Project, focusing on ultra-short-period contact binaries, among which V1104 Her was included. Their photometric observations were carried out over 40 nights between 2016 and 2021 using $BVRI$ filters. Light curve analysis presented in their study yielded the parameters $T_{1}=4050$ K, $T_{2}=3833(24)$ K, $q=0.967(16)$, $f=0.16(5)$, and $i=82.3(2)^\circ$, while no stellar spot was required to reproduce the observed light curves. Temperatures derived in the present study are approximately 300 K higher than those reported by \cite{loukaidou2022cobitom}, primarily because the initial temperature adopted here was constrained using Gaia DR3 data, which indicate a temperature near 4200 K. The largest discrepancy between the two investigations is related to the measured mass ratio, since the solutions imply different identifications for the more massive component. Part of this inconsistency may originate from the temporal separation between the observational datasets, considering that contact binary systems undergo continuous evolutionary changes. Fillout factors obtained in both analyses consistently indicate that V1104 Her is a shallow-contact system. Absence of any requirement for starspot modeling in either solution further suggests that the system has remained relatively stable against significant magnetic activity over the investigated time interval. Absolute parameters reported by \cite{loukaidou2022cobitom} are $M_{1}=0.824(17)\,M_{\odot}$, $M_{2}=0.797(19)\,M_{\odot}$, $R_{1}=0.764(5)\,R_{\odot}$, $R_{2}=0.701(5)\,R_{\odot}$, $L_{1}=0.141(28)\,L_{\odot}$, and $L_{2}=0.095(20)\,L_{\odot}$. Differences regarding the identification of the more massive component are also reflected in the derived absolute parameters in the present study, although the overall ranges of the physical quantities remain broadly comparable between the two investigations.

Based on the analyses presented by \cite{2015AJ....149..148L} and \cite{loukaidou2022cobitom}, the orbital period behavior of V1104 Her exhibits a complex structure that cannot be adequately described by a simple secular trend alone. \cite{2015AJ....149..148L}'s results indicate that, in addition to a long-term decrease in the orbital period, the system shows pronounced cyclic variations. In the \cite{2015AJ....149..148L} study, a combined model consisting of a quadratic term and two periodic components was found to provide the best representation of the observed O-C behavior. The quadratic term was interpreted by \cite{2015AJ....149..148L} as evidence for a continuous decrease in the orbital period, consistent with angular momentum loss and/or mass transfer processes commonly expected in contact binary systems. The superimposed cyclic variations were attributed to the light-travel time effect, suggesting the presence of additional companions gravitationally bound to the eclipsing pair. Alternative explanations, such as magnetically driven mechanisms, were considered less plausible by \cite{2015AJ....149..148L} due to physical inconsistencies with the required stellar parameters. On this basis, \cite{2015AJ....149..148L} proposed that V1104 Her is likely a higher-order multiple system, potentially hosting more than one additional body. \cite{loukaidou2022cobitom} also indicated the existence of a third body based on a cyclic trend in their analysis. Therefore, they calculated $P_3\,(\mathrm{yr}) = 8.35(19)$, $A\,(\mathrm{day}) = 0.0012(1)$, $M_{3,\mathrm{min}}\,(M_\odot) = 0.11(1)$, and $a_{12}\sin i_3\,(\mathrm{au}) = 0.324(16)$. Based on our analysis of V1104 Her, incorporating the newly derived TESS eclipse timings together with the available ground-based measurements, we confirm that the O-C diagram exhibits a complex structure with evidence for cyclic modulation. A similar behavior is also found for V1284 Her. Statistical model comparison using the AIC, BIC, and $\chi^2$ criteria indicates that the cyclic model provides a significantly better description of the currently available eclipse timings for both V1104 Her and V1284 Her than the linear ephemeris. For V2822 Ori, the comparison is less conclusive: although the cyclic model yields a slightly lower AIC value, the BIC favors the simpler linear model because of its smaller number of free parameters. Nevertheless, given the small difference between the two models, the sinusoidal solution was also examined for consistency and comparison with the other systems. If the observed modulations are interpreted as the light-travel time effect, the derived orbital parameters suggest the presence of candidate tertiary companions. These systems are expected to constitute promising targets for future investigations of orbital period variations in close binary systems.

\begin{figure*}
\centering
\includegraphics[width=0.99\textwidth]{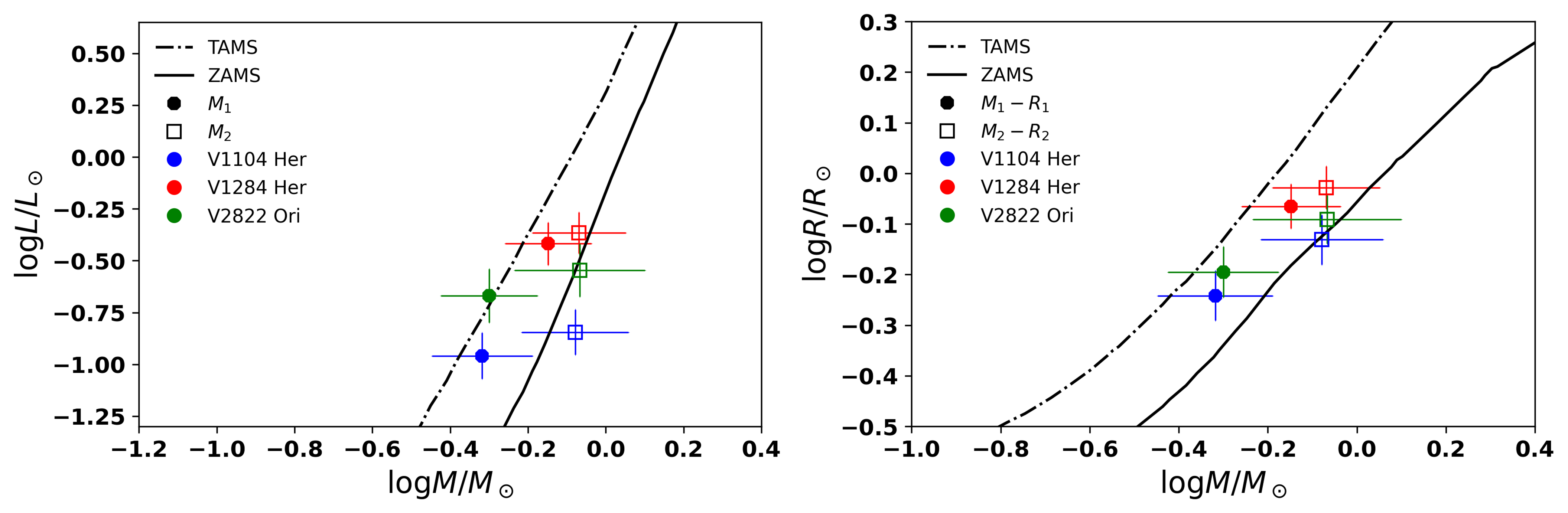}
\caption{$M-L$ and $M-R$ diagrams for the studied contact binaries. The systems are displayed in different colors.}
\label{Fig:MLR}
\end{figure*}

According to the estimated absolute parameters (Table \ref{tab:absolute}), Mass–Luminosity ($M$–$L$) and Mass–Radius ($M$–$R$) diagrams in logarithmic form are employed to investigate the evolutionary status of the target systems (Figure \ref{Fig:MLR}). The stellar components are shown relative to the Zero-Age Main Sequence (ZAMS) and Terminal-Age Main Sequence (TAMS) boundaries defined by \cite{2000A&AS..141..371G}. In $M$–$L$ and $M$–$R$ diagrams, the less massive companions are expected to appear closer to the TAMS, whereas the more massive stars are anticipated to lie nearer to the ZAMS. Contact binaries follow complex evolutionary paths due to mass and angular momentum transfer (\citealt{2005ApJ...629.1055Y}). Their positions relative to the ZAMS and TAMS boundaries derived from evolutionary models for single stars should therefore be approached with appropriate caution.

\vspace{0.6cm}
\section*{Data Availability}
The ground-based observations and times of minima used in this study are available in the online supplementary materials of the paper.

\vspace{0.6cm}
\section*{Acknowledgments}
This work, including the observations, analyses, and writing, was conducted under the BSN project. The analysis of two of the systems is based on observations carried out at the Observatorio Astron\'omico Nacional on the Sierra San Pedro M\'artir, operated by the Universidad Nacional Aut\'onoma de M\'exico. Data reduction was performed using IRAF, distributed by the National Optical Observatories and operated by the Association of Universities for Research in Astronomy, Inc., under a cooperative agreement with the National Science Foundation. This study also makes use of results from the European Space Agency's Gaia mission (\url{http://www.cosmos.esa.int/gaia}) and of observations obtained by the TESS mission, supported through NASA's Explorer Program. We are deeply grateful to Ehsan Paki for his valuable assistance.

\vspace{0.6cm}
\bibliography{References}{}
\bibliographystyle{aasjournal}
\end{document}